\documentclass{elsart}
\usepackage{amsmath}
\usepackage{amssymb}
\usepackage{epsfig}

\begin{document}
\begin{frontmatter}
\title      {Configuration mixing of mean-field wave-functions projected
             on angular momentum and particle number; application to
$^{24}$Mg}

\author     {A. Valor, P.-H. Heenen}
\address {Service de Physique Nucl\'{e}aire Th\'{e}orique et de
Physique Math\'{e}matique,
 U.L.B - C.P. 229, B 1050 Brussels, Belgium}
\author     {P. Bonche}
\address { SPhT CEA Saclay, F 91191 Gif sur Yvette Cedex, France}

\date{\today}

\maketitle


\begin{abstract}
We present in this paper the general framework
of a method which permits to restore the rotational
and particle number symmetries of wave functions obtained in
Skyrme HF+BCS calculations. 
This restoration is nothing but a projection of mean-field intrinsic
wave functions onto good particle number and good angular momentum.
The method allows also to {\it mix} projected wave functions. 
Such a configuration mixing is discussed for sets of HF+BCS intrinsic states
generated in constrained calculations with suitable collective
variables.
This procedure gives collective states which are eigenstates of
the particle number and the angular momentum operators and between
which transition probabilities are calculated.
An application to $^{24}$Mg is presented, with mean-field wave functions
generated by axial quadrupole constraints.
Theoretical spectra and
transition probabilities are compared to the experiment.

PACS numbers: 21.10.Ky 21.30.-n 21.60.Jz 27.30+t

\end{abstract}
\end{frontmatter}

\section{Introduction}

The cranking method is widely used in nuclear
spectroscopy to describe high spin states. 
Applications based on effective nuclear interactions have been
particularly successful in the description of super-deformed 
rotational bands in several regions of the mass 
table~\cite{Gal94,Ter95,Rig99,Rob98,Val99}. 
In the cranking method, a rotational band is generated by the 
rotation of a deformed intrinsic state. 
Since cranking states are not eigenstates of angular momentum,
it is not straightforward to determine transition rates
in nuclei which are not very well deformed. 
To overcome this difficulty, 
approximations~\cite{Isl79,Man75,RS80} 
have been developed for transitions  within a band. 
However, they are only valid when the structure of the nuclear 
states are not affected by rotation, i.e. for rigid nuclei. 

Another limitation of the cranking model occurs in nuclei soft 
with respect to the variation of a collective variable. 
In this case, one expects that the interference of the zero-point 
vibrational mode with the rotational motion will lead to 
variations in the nuclear structure along the yrast line.

There have been several attempts in the past to introduce
phenomenological corrections taking into account rotational
and vibrational corrections to the mean-field energy (see for
instance~\cite{Cam74,Gir79,Rei99}).
It is the aim of the present paper to introduce a method in which 
rotations and vibrations are taken into account simultaneously 
in a more general and consistent way. 

The starting point of our approach is a set of many-body 
wave functions generated by constrained Skyrme HF+BCS calculations.
The discretization of these wave functions 
on a 3-dimensional Cartesian mesh enables to describe very general 
shapes of the nuclear density and to easily write the effect of a
spatial rotation on the mean-field wave functions.
This property permits to restore symmetries with respect to angular 
momentum and to the proton and neutron particle numbers
in a systematic way.

It is not the first time that symmetries are restored
on mean-field wave functions. 
Several methods have been developed for schematic nuclear interactions 
and a limited number of active valence nucleons (for recent references,
see~\cite{Har95,Ena99}).
Methods based on effective interactions have been limited to either 
angular momentum~\cite{Cau77,Bay84,Rod99} or particle number
projection with mixing of configurations by
the generator coordinate method (GCM)~\cite{Hee93}.
The present work generalizes  this last work by the inclusion of 
a projection on angular momentum.

We present below the general framework of our method together with 
a test on a light nucleus for which extensive calculations 
can be performed.
In the first part, we show how to implement the projection 
on $\vec J$, $N$ and $Z$ simultaneously and to determine 
the contractions needed to calculate the matrix elements between 
two different many-body HFB vacua. 
The specificities of mesh calculations require to take explicitly into 
account the fact that the single-particle bases are not complete  and lead to
formulae rather different from some similar works~\cite{Egi93,Ena99}.
In the second part of this work, we present an application to the 
$^{24}$Mg nucleus for which the properties of several effective 
interactions are
tested.

\section{Angular momentum and particle number projections }

\subsection{Principle of the method}

The starting point of our method is a set of wave functions 
$ |\Phi{_\alpha}\rangle$ generated by mean-field calculations
with a constraint on a collective coordinate~${\alpha}$.
Wave functions with good angular momentum and particle numbers are 
obtained by restorations of symmetry on~$ |\Phi_{\alpha}\rangle$:
\begin{equation}
|\Phi,JM\alpha\rangle = \frac{1}{\mathcal N}\sum_{K}g_K
 {\hat{P}^{J}_{MK}\hat{P}^{Z}\hat{P}^{N} |\Phi_{\alpha}\rangle}\quad ,
\end{equation}
where $\mathcal N$ is a normalization factor, 
and the operators $\hat{P}$ are projectors.

A configuration mixing on the collective variable $\alpha$
is then performed for each angular momentum:
\begin{equation}{\label{discsum}}
|\Psi,JM\rangle =\sum_{\alpha}f_{\alpha}^{JM}|
                              \Phi,JM\alpha\rangle \quad .
\end{equation}
The weight functions $f_{\alpha}^{JM}$ are found by requiring
that the expectation value of the energy:
\begin{equation}\label{E10a}
E^{JM}={\langle{\Psi, JM}\vert\hat H\vert{\Psi, JM}\rangle
    \over\langle{\Psi, JM}\vert{\Psi, JM}\rangle} \quad ,
\end{equation}
is stationary with respect to an arbitrary  variation 
$\delta f_{\alpha}^{JM}$.
This prescription leads to the discretized Hill-Wheeler
equation~\cite{HW53}:
\begin{equation}\label{E11}
 \sum_{\alpha}({\mathcal H}^{JM}_{\alpha,\alpha'}
      -E^{JM}_k{\mathcal I}^{JM}_{\alpha,\alpha'})
          f^{JM,k}_{\alpha'}=0\quad,
\end{equation}
in which the Hamiltonian kernel ${\mathcal H}^{JM}$ and the overlap kernel
${\mathcal I}^{JM}$ are defined as
\begin{equation}\label{E12}
{\mathcal H}^{JM}_{\alpha,\alpha'}=
     \langle{\Phi JM\alpha}\vert\hat H\vert{\Phi JM\alpha'}\rangle,\quad
{\mathcal I}^{JM}_{\alpha,\alpha'}=
     \langle{\Phi JM\alpha}\vert{\Phi JM\alpha'}\rangle\quad.
\end{equation}
Since the Hamiltonian is rotationally invariant (we will come back 
to the problem of the density dependence of the interaction 
in section~\ref{ddofsf}) and conserves the number of particles, 
one has  to restore the symmetries on only one of the two wave functions
entering in each matrix element like eq.~(\ref{E12}).
The kernels are obtained by integration on three
Euler angles and two gauge angles of the matrix elements between
rotated wave functions.

Besides these kernels, we will calculate transition probabilities
between different eigenstates of the Hill-Wheeler equation.
This requires the calculation of the matrix elements of a tensor 
of order L, $\hat{T}^{M}_{L}$, between projected states.

Such a secular problem based on the configuration mixing defined
by eq.~\ref{discsum} amounts to a variation after projection in
a many-body Hilbert space built on a limited set of states obtained 
for different values of the collective variables~$\alpha$'s.

\subsection{Determination of the contractions}{\label{detcon}}

The matrix elements (eq.~\ref{E12}) are calculated in two steps. 
They are first determined for a given rotation and for given 
gauge angles. 
Then they are integrated over the Euler and gauge angles.

To compute the kernels, we make use of the Balian-Br\'ezin 
theorem~\cite{BB69} which expresses them in terms of contractions 
of creation and/or annihilation single-particle operators.

To simplify the presentation, the spatial rotation is applied 
on the left wave function, whereas the particle number rotation 
is applied on the right one. 
The wave functions after rotation are denoted by $|L\rangle$  
and $|R\rangle$ and the associated quasiparticle  annihilation
operators by $l_{\mu}$, $r_{\mu}$ respectively, with $\mu=1,\dots,N$. 
The number N is the total number of HF states included 
in the mean-field calculations.

The quasiparticle operators are obtained by the diagonalization
of the Bogoliubov equations on a single-particle basis.
When the equations are discretised on a 3D mesh, a convenient basis 
is provided by the eigenstates of the HF Hamiltonian. 
Since only a limited number of single-particle HF states are 
determined, the left and right bases are different.
The particle annihilation and  creation operators of the left and
right particle bases will be denoted $(a,a^{+})$ and  $(b,b^{+})$  
respectively.
They are related by a unitary transformation:
\begin{equation}
a_{\mu}=\sum_{\nu}  R_{\mu,\nu} b_{\nu} \quad .
\end{equation}
The quasiparticle operators $l_{\mu}$ and $r_{\mu}$ are expressed 
as a function of the left and right single-particle bases
by Bogoliubov transformations~\cite{RS80}:
\begin{gather}
\begin{pmatrix}
{\mathbf{l}} \\
{\mathbf{l}}^{+}
\end{pmatrix}
=
\begin{pmatrix}
{U_{l}^{B}}^{\dag} & {V_{l}^{B}}^{\dag}\\
{V_{l}^{B}}^T      & {U_{l}^{B}}^T
\end{pmatrix}
\begin{pmatrix}
{\mathbf{a}} \\
{\mathbf{a}}^{+}
\end{pmatrix}
\equiv W^{\dag}
\begin{pmatrix}
{\mathbf{a}} \\
{\mathbf{a}}^{+}
\end{pmatrix}\quad ;
\end{gather}
where the matrices $U_{l}^{B}$ and $V_{l}^{B}$ have  dimensions
N x N. The expressions for the right quasi-particles are similar.

Using the same notations as in~\cite{BDF90}, we  define the two 
vectors
\begin{equation}
\lambda=
\begin{pmatrix}
{\mathbf{l}} \\{\mathbf{l}^{+}}
\end{pmatrix}\quad , \qquad
\rho=
\begin{pmatrix}
{\mathbf{r}} \\{\mathbf{r}^{+}}
\end{pmatrix} \quad , \qquad
\end{equation}
in terms of which any two-body operator $\hat{O}$ can be expressed.

The quasiparticle bases are related by a linear transformation 
$\hat{T}$. 
Taking into account that the matrix~$R$ linking the HF single-particle 
bases does not mix creation and annihilation operators, one obtains:
\begin{equation}
\lambda=\hat{T}\rho, \;\;\; \hat{T}=
\begin{pmatrix}
(D^{-1})^{*} & -E \\
-E^{*} & D^{-1}
\end{pmatrix},
\end{equation}
where,
\begin{equation}
D=({U_{l}^{B}}^{T}R{U_{r}^{B}}^{*} + {V_{l}^{B}}^{T}R{V_{r}^{B}}^{*})^{-1},
\;\;\;\;\;  E=-({U_{l}^{B}}^{+}R{V_{r}^{B}}^{*} + {V_{l}^{B}}^{+}R{U_{r}^{B}}^{*}),
\end{equation}

Balian and Br\'ezin~\cite{BB69} have established the relation
between $\hat{T}$ and the contractions $\langle\lambda\rho
\rangle$, $\langle\rho\rho\rangle$ and $\langle\lambda\lambda\rangle$:
\begin{equation}
\langle\lambda\lambda\rangle=
\begin{pmatrix}
\langle{\mathbf{l}}{\mathbf{l}}\rangle & 1\\
0                    & 0
\end{pmatrix}
=
\begin{pmatrix}
ED & 1\\
0                    & 0
\end{pmatrix} \quad ,
\end{equation}
\begin{equation}
\langle\rho\rho\rangle=
\begin{pmatrix}
0 & 1\\
0 & \langle{\mathbf{r}}^{+}{\mathbf{r}}^{+}\rangle
\end{pmatrix}
=
\begin{pmatrix}
0 & 1\\
0                    & DE^{*}
\end{pmatrix} \quad ,
\end{equation}
\begin{equation}
\langle\lambda\rho\rangle=
\begin{pmatrix}
0 & \langle {\mathbf{l}}{\mathbf{r}}^{+} \rangle\\
0 & 0
\end{pmatrix}
=
\begin{pmatrix}
0 & D^{T}\\
0 & 0
\end{pmatrix} \quad .
\end{equation}
In addition, they have derived a formula for the modulus of
the overlap between left and right many-body wave functions:
\begin{equation}\label{ovbalbre}
\langle L|R\rangle = \pm(\det D^{-1})^{1/2}
\end{equation}

One obtains the contractions in the particle bases:
\begin{eqnarray}
\langle {\mathbf{a}}^{+}{\mathbf{b}} \rangle&=&{V_{l}^{B}}D^{T}{V_{r}^{B}}^{+}
\quad ,\nonumber \\
\langle {\mathbf{a}}^{+}{\mathbf{a}}^{+} \rangle&=&{V_{l}^{B}}{U_{l}^{B}}^{+} +
{V_{l}^{B}}ED{V_{l}^{B}}^{T} \quad ,  \\
\langle {\mathbf{b}}{\mathbf{b}} \rangle&=&{U_{r}^{B}}{V_{r}^{B}}^{+} +
{V_{r}^{B}}^{*}DE^{*}{V_{r}^{B}}^{+} \quad .  \nonumber
\end{eqnarray}

\subsection{Symmetry restrictions}{\label{symres}}

The formulae derived in section \ref{detcon} are general.
In the present application, we impose several symmetry 
restrictions on the mean-field wave functions.
These restrictions which we develop below do not limit the 
formalism, they are only intended to simplify the equations 
for our test case.
They can be released, if need be, at the only cost of additional 
computational time.

First, the total wave function is symmetric with respect to reflections 
across the $x=0$,  $y=0$ and $z=0$ planes and time-reversal invariant.
With this symmetry restriction to ellipsoidal configurations, the
single-particle wave functions are described by
4-dimensional vectors~\cite{BFH87},
\begin{equation}
\begin{pmatrix}
w_{1,\mu} \\ w_{2,\mu} \\ w_{3,\mu} \\ w_{4,\mu}
\end{pmatrix}
=
\begin{pmatrix}
Re \; \Phi_{\mu}({\bf r},+,+,+p) \\
Im \; \Phi_{\mu}({\bf r},-,-,+p) \\
Re \; \Phi_{\mu}({\bf r},-,+,-p) \\
Im \; \Phi_{\mu}({\bf r},+,-,-p)
\end{pmatrix}\quad ,
\end{equation}
where the components describe, respectively, the real and imaginary
spin up and down parts of the single-particle wave functions.
The total parity is given by p.
Each of the components has a well defined symmetry
with respect to reflections across the three planes  $x=0$, $y=0$ 
and $z=0$ (see Ref.~\cite{BDF90}, appendix C).
In a mesh calculation, the space is limited to a box beyond which 
all the wave functions are set to zero.
With these symmetries, all the integrations can be limited to 
an octant of the box.

A second restriction is that the intrinsic wave functions have been 
generated by constraints on the axial quadrupole moment only,
the z-axis being the symmetry axis of the nucleus.
To restore angular momentum requires then only to rotate
the mean-field wave function by an angle $\beta$ around the
$y$-axis. The  coordinates $(x_{0},y_{0},z_{0})$
become after rotation:
\begin{eqnarray}
x_{1}&=&z_{0}\sin(\beta) + x_{0} \cos(\beta) \quad , \nonumber \\
y_{1}&=&y_{0} \quad , \\
z_{1}&=&z_{0}\cos(\beta) - x_{0} \sin(\beta) \quad ,  \nonumber
\end{eqnarray}
and the four components  of a wave function
rotate according to:
\begin{eqnarray}\label{paritywf}
p_{1,\mu}({\bf r}_{1}) &=& w_{1,\mu}({\bf r}_{0}) \cos(\beta/2) -
w_{3,\mu}({\bf r}_{0}) \sin(\beta/2) \nonumber \\
p_{2,\mu}({\bf r}_{1}) &=& w_{2,\mu}({\bf r}_{0}) \cos(\beta/2) -
w_{4,\mu}({\bf r}_{0}) \sin(\beta/2) \nonumber \\
p_{3,\mu}({\bf r}_{1}) &=& w_{3,\mu}({\bf r}_{0}) \cos(\beta/2) +
w_{1,\mu}({\bf r}_{0}) \sin(\beta/2) \nonumber \\
p_{4,\mu}({\bf r}_{1}) &=& w_{4,\mu}({\bf r}_{0}) \cos(\beta/2) +
w_{2,\mu}({\bf r}_{0}) \sin(\beta/2)\quad .
\end{eqnarray}

These equations show that while the total parity p is  preserved
by a rotation, the specific symmetries of the four components with respect
to the $x=0$ and $z=0$ planes are lost. In the same way, signature
as defined in ref~\cite{BFH87} is not conserved. An immediate
consequence of this loss of symmetry is that matrix
elements between rotated and non rotated wave functions must
be calculated in half of a box. The real
and imaginary parts of the rotated wave function have different
symmetries with respect to the $y=0$ plane and the overlap
between rotated and non rotated individual wave functions are real.

Furthermore, the points $(x_{1},y_{1},z_{1})$
do not coincide after rotation with mesh points.
The values of the rotated wave functions
on the mesh points are reevaluated using the analytical forms of
functions defined on a mesh given in ref~\cite{BH86}.

The relation between rotated (left) and non rotated
(right) operators is given by:
\begin{equation}
a_{\mu}=\sum_{\nu>0} \{ R_{\mu,\nu} b_{\nu} +
R_{\mu,\overline{\nu}} b_{\overline{\nu}} \}\quad , \qquad
a_{\overline{\mu}} = \sum_{\nu>0} \{ R_{\overline{\mu} ,\nu} b_{\nu}+
R_{\overline{\mu},\overline{\nu}} b_{\overline{\nu}} \}\quad ,
\end{equation}
where the state $|\overline{\mu}\rangle$ denotes the time-reversed
partner of state $|\mu\rangle$ and the matrix $R$ is given by the
overlap between left and right states.
Using the properties of the time reversal operator, one
can show the following relations between the overlaps:
\begin{equation}
R_{\overline{\mu},\nu}= -R_{\mu,\overline{\nu}} \quad , \qquad
R_{\overline{\mu},\overline{\nu}}=R_{\mu,\nu} \quad .
\label{Rform}
\end{equation}
The matrix R does not separate into blocks corresponding
to the different combinations of parity and signature
as in our previous study~\cite{Hee93}.

\subsection{Restriction to BCS transformations}{\label{symbcs}}

One can still simplify the expressions for contractions derived in
sub-section~\ref{detcon} for the general Bogoliubov transformations 
to BCS transformations.
This limitation should not be too bad for even nuclei close to
the stability line, as $^{24}$Mg, for which pairing correlations
do not couple significantly bound and continuum states.

The state $|R\rangle$ is obtained from the original intrinsic state 
by multiplying by a phase $e^{2i\phi}$ the occupation numbers in the 
original BCS (or HFB) transformation
\begin{equation}
|R\rangle \equiv |R(\phi)\rangle   \Longleftrightarrow
\left\{
\begin{array}
rr_{\mu}  = u_{r\mu}b_{\mu} -v_{r\mu}e^{i2\phi}b_{\overline{\mu}}^{+}  \nonumber\\
r_{\overline{\mu}}  = u_{r\mu}b_{\overline{\mu}} +
v_{r\mu}e^{i2\phi}b_{\mu}^{+} \nonumber
\end{array} \right.,
\end{equation}
with $u_{r\mu}$, $v_{r\mu}$ real and positive, and $u_{r\overline{\mu}} =
u_{r\mu}$, $v_{r\overline{\mu}} = -v_{r\mu}$.

In the BCS case the matrices ${U_{l}^{B}}$,
${V_{l}^{B}}$, ${U_{r}^{B}}$, ${V_{r}^{B}}$ take the form
\begin{equation}
{U_{l}^{B}}\equiv{U_{l}}\; ; \;\;\; {V_{l}^{B}}\equiv -\sigma {V_{l}}\; ; \;\;\;
{U_{l}}=\begin{pmatrix}
u_{l} & 0 \\
0     & u_{l}
\end{pmatrix}\; ; \;\;\;
 {V_{l}}=\begin{pmatrix}
v_{l}        & 0 \\
0   & v_{l}
\end{pmatrix}\; ;
\end{equation}
\begin{equation}
{U_{r}^{B}}\equiv{U_{r}}\; ; \;\;\; {V_{r}^{B}}\equiv -\sigma {V_{r}}e^{-i2\phi}
\; ; \;\;\; {U_{r}}=\begin{pmatrix}
u_{r} & 0 \\
0     & u_{r}
\end{pmatrix}\; ; \;\;\;
 {V_{r}}=\begin{pmatrix}
v_{r}   & 0 \\
0       & v_{r}
\end{pmatrix}\; ;
\end{equation}

The matrices $u_{l}$, $v_{l}$, $u_{r}$, $v_{r}$, are real, diagonal and of
dimension N/2 x N/2. The matrix $\sigma$ is defined as
\begin{equation}
\sigma=
\begin{pmatrix}
0 & -1 \\
1 & 0
\end{pmatrix}\quad .
\end{equation}

Thanks to the symmetry properties given by equation~(\ref{Rform}), the
matrices D and E introduced in subsection~\ref{detcon} have the same
block structure as matrix R.
These same symmetries allow the relation $\sigma^{T} R \sigma = R$.
We can thus write the contractions, as well as matrices D, E, in a
simpler way. We have:
\begin{xxalignat}{2}
\langle {\mathbf{a}}^{+}{\mathbf{b}} \rangle & = V_{l} D^{T} V_{r}
e^{2i\phi}\; ;  & \quad D & =V_{l}RV_{r}e^{2i\phi} + U_{l} R U_{r}\; ; \quad \\
\quad \langle {\mathbf{a}}^{+}{\mathbf{a}}^{+} \rangle & = -\sigma ( V_{l}
U_{l} - V_{l} \tilde{E}DV_{l} )\; ; &  \quad E & = \sigma \tilde{E}\; ; \\
\langle {\mathbf{b}}{\mathbf{b}} \rangle & = \sigma ( U_{r}
V_{r} + V_{r} D\tilde{E}^{*}V_{r} e^{2i\phi})e^{2i\phi}\; ; & \qquad
\tilde{E} & = U_{l} R V_{r} e^{2i\phi} - V_{l}RU_{r}\; .
\end{xxalignat}

\subsection{Elimination of non occupied states}

In the derivation shown in the above section, it is assumed 
that left and right bases are either complete
or truncated in such a way that they span the same space.
As discussed in ref~\cite{BDF90}, this property is not valid when
orbitals are discretized on a three dimensional mesh.
This problem can be solved by taking into account that the missing 
part of the expansion of the left states on the right basis includes
empty states that do not affect the structure of the nucleus. 
These states are defined by the condition $v_{\mu}=0$ and
contribute neither to the overlap nor to the contractions.
Since the structure of the matrices is different
than in ref~\cite{BDF90}, we give now the formulae corresponding to 
the present case.

Let us introduce:
\begin{gather}
{{V}_{l}}
= \begin{pmatrix} \overline{{V}}_{l} & 0 \\ 0 & 0 \end{pmatrix}\; 
; \;\;\;
{{U}_{l}}
= \begin{pmatrix} \overline{{U}}_{l} & 0 \\ 0& 1 \end{pmatrix}\; ;
\label{reorga}
\end{gather}
and split the unitary matrix R in the form
\begin{eqnarray}
R=
\begin{pmatrix} {\mathcal{R}} &{\mathcal{I}}\\
{\mathcal{T}}&{\mathcal{U}} \end{pmatrix}\; .
\end{eqnarray}

With these notations, the contractions can be written as:
\begin{eqnarray}\label{cona+b}
\langle \mathbf{a}^{+} \mathbf{b} \rangle =
\begin{pmatrix}
{\overline{{V}}}_{l}{\mathcal{D}}^{T} {\overline{{V}}}_{r}e^{2i\phi} & 0\\
0 & 0 \end{pmatrix}\quad ;
\end{eqnarray}
\begin{eqnarray}
\langle \mathbf{b}\mathbf{b} \rangle =
\sigma
\begin{pmatrix}
\overline{{U}}_{r}\overline{{V}}_{r} +
\overline{{V}}_{r}{\mathcal{D}}\mathcal{E}_{r} \overline{{V}}_{r} & 0 \\ 0 & 0
\end{pmatrix}\quad ;
\end{eqnarray}
\begin{eqnarray}
\langle \mathbf{a}^{+}\mathbf{a}^{+}\rangle = -\sigma \begin{pmatrix}
\overline{{V}}_{l}\overline{{U}}_{l} -
\overline{{V}}_{l}\mathcal{E}_{l}{\mathcal{D}} \overline{{V}}_{l} & 0 \\0 &0
\end{pmatrix}\quad .
\end{eqnarray}
The matrices $\mathcal{D}$, $\mathcal{E}_{r}$ and $\mathcal{E}_{l}$ are defined by:
\begin{eqnarray}\label{matd}
{\mathcal{D}} = \left({\overline{{U}}}_{l} ({\mathcal{R}}^{+})^{-1}
{\overline{{U}}}_{r} + {\overline{{V}}}_{l}
{\mathcal{R}} {\overline{{V}}}_{r} e^{2i\phi} \right)^{-1}\quad ;
\end{eqnarray}
\begin{eqnarray}
\mathcal{E}_{r} = \overline{{U}}_{l} ({\mathcal{R}}^{+})^{-1}
\overline{{V}}_{r}  - \overline{{V}}_{l}{\mathcal{R}}\overline{{U}}_{r}e^{2i\phi}\quad ;
\end{eqnarray}
\begin{eqnarray}
\mathcal{E}_{l} = \overline{{U}}_{l} {\mathcal{R}}
\overline{{V}}_{r}e^{2i\phi}  - \overline{{V}}_{l}({\mathcal{R}}^{+})^{-1}
\overline{{U}}_{r}\quad .
\end{eqnarray}

\subsection{Determination of the overlap}

\subsubsection{Modulus}

Equation~(\ref{ovbalbre}) gives the overlaps up to a sign.
Several methods have been developed to calculate the global 
phase of the overlap. 
We show in the next section which procedure we have implemented.
In this section, we give the explicit formulae for the modulus 
of the overlap.
We have:
\begin{eqnarray}
D^{-1} = {U}_{l} R {U}_{r} + {V}_{l} R {V}_{r}e^{2i\phi}\quad ,
\end{eqnarray}
so that:
\begin{eqnarray}
\langle L | R\rangle&=&\pm ( \det (D^{-1}) )^{1/2} \nonumber \\
&=& \pm \left( \det
\begin{pmatrix}
\overline{{U}}_{l}{\mathcal{R}}\overline{{U}}_{r} +
\overline{{V}}_{l}{\mathcal{R}}\overline{{V}}_{r}e^{2i\phi} &
\overline{{U}}_{l} {\mathcal{I}} \\
{\mathcal{T}} \overline{{U}}_{r} & {\mathcal{U}}
\end{pmatrix}
\right)^{1/2}\quad .
\end{eqnarray}
It is easy, although tedious, to demonstrate that the matrix $D$,
\begin{eqnarray}
D= \begin{pmatrix}
{\mathcal{D}}; & {\mathcal{D}} \overline{{U}}_{l}
({\mathcal{R}}^{+})^{-1} {\mathcal{T}}^{+} \\
{\mathcal{I}}^{+}({\mathcal{R}}^{+})^{-1} \overline{{U}}_{r} {\mathcal{D}}; &
{\mathcal{U}}^{-1} + {\mathcal{I}}^{+} ({\mathcal{R}}^{+})^{-1} \overline{{U}}_{r}
{\mathcal{D}} \overline{{U}}_{l}  ({\mathcal{R}}^{+})^{-1} {\mathcal{T}}^{+}
\end{pmatrix}\; ,
\end{eqnarray}
where $\mathcal{D}$ is given by eq.~\ref{matd},
can be expressed as the product of four matrices~\cite{FLO75}:
\begin{eqnarray}
D =
\begin{pmatrix} \overline{{U}}_{r}^{-1} & 0 \\ 0 & 1 \end{pmatrix}
\begin{pmatrix} {\mathcal{R}}^{+} &{\mathcal{T}}^{+}  \\
{\mathcal{I}}^{+} & {\mathcal{U}}^{+}  \end{pmatrix}
\begin{pmatrix} \alpha_{r}^{-1} & 0 \\ 0 & 1 \end{pmatrix}
\begin{pmatrix} 1 & -\beta_{r} \\ 0 & 1 \end{pmatrix}\quad ,
\end{eqnarray}
with:
\begin{eqnarray}
&&\alpha_{r} = \overline{{U}}_{l} + \overline{{V}}_{l} {\mathcal{R}}
\overline{{V}}_{r} \overline{{U}}_{r}^{-1} {\mathcal{R}}^{+}e^{2i\phi}\; , \nonumber \\
&&\beta_{r} = \overline{{V}}_{l} {\mathcal{R}} \overline{{V}}_{r}
\overline{{U}}_{r}^{-1} {\mathcal{T}}^{+}e^{2i\phi}\; ;
\end{eqnarray}
so that
\begin{eqnarray}
\det D &=& \det ( \overline{{U}}_{r}^{-1} ) \det ( {{R}}^{+} )
\det ( \alpha_{r}^{-1} ) \nonumber \\
&=& \det ( \overline{{U}}_{r}^{-1} \alpha_{r}^{-1})
 \nonumber \\
&=&\det \left[ ( ({\mathcal{R}}^{+})^{-1} ) \right] \det ({\mathcal{D}})\; ,
\end{eqnarray}
where we have taken into account that the determinant of the unitary
matrix R equals 1.
We finally obtain:
\begin{eqnarray}
\langle L | R \rangle(\phi) = \pm \left( \frac{ \det {\mathcal{R}} }{ \det
{\mathcal{D}} } \right)^{1/2}\quad .
\end{eqnarray}

\subsubsection{Phase}

For any space rotation angle $\beta$, the overlap between two 
mean-field wave functions is positive for $\phi_n=0$ and $\phi_p=0$.
We need to follow the evolution of the phase of the overlap along 
the integration paths on $\phi_n$ and $\phi_p$.
One has:
\begin{eqnarray}\label{elover}
\langle L|R\rangle(\phi) =
\sqrt{ (\text{Re} [\langle L|R\rangle(\phi)])^{2} + (\text{Im}
[\langle L|R\rangle(\phi)])^{2} }
e^{i\Omega(\phi) }\; .
\end{eqnarray}
We can rewrite the global phase $\Omega$ as:
\begin{eqnarray}
\Omega(\phi) = \text{Im} \ln ( \langle L|R\rangle(\phi) )
 = \Phi(\phi) +n(\phi)\pi\; ,
\end{eqnarray}
where $\Phi$, which can be defined as:
\begin{eqnarray}\label{phasedolia}
\Phi(\phi) = \arctan ( \text{Im} [\langle L|R\rangle(\phi)]/
\text{Re} [\langle L|R\rangle(\phi)] )\quad ,
\end{eqnarray}
is limited to the interval $[-\pi/2\;,\,\pi/2]$.
The integer number $n$ remains to be determined.
The total phase and its derivative are given by:
\begin{eqnarray}\label{phasder}
\frac{d\Omega(\phi)}{d\phi} =
\text{Re} \left\{ \frac{\langle L |
\hat{N}  |R\rangle(\phi)}
{\langle L| R\rangle(\phi) } \right\}\quad .
\end{eqnarray}
It can also be calculated directly from the matrix element 
of the particle number operator:
\begin{eqnarray}
\frac{\langle L | \hat{N}  |R\rangle(\phi)} {\langle L| R\rangle(\phi) }
&=& \sum_{\mu>0} \langle a^{+}_{\mu}
a_{\mu} \rangle_{\phi} + \langle a^{+}_{\overline{\mu}}
a_{\overline{\mu}} \rangle_{\phi} \; , \nonumber \\
&=&2 \sum_{\mu,\mu'>0} \{ R_{\mu\mu'} \langle a^{+}_{\mu}
b_{\mu'} \rangle_{\phi} + R_{\mu\overline{\mu}'} \langle a^{+}_{\mu}
b_{\overline{\mu}'} \rangle_{\phi} \} \; , \\
&=&2 \sum_{\mu,\mu'>0} \{ {\mathcal{R}}_{\mu\mu'} (\overline{{V}}_{l}
{\mathcal{D}}^{T} \overline{{V}}_{r} )_{\mu\mu'} \text{e}^{2i\phi} +
{\mathcal{R}}_{\mu\overline{\mu}'} (\overline{{V}}_{l}
{\mathcal{D}}^{T} \overline{{V}}_{r} )_{\mu\overline{\mu}'} \text{e}^{2i\phi} \}\; , \nonumber
\end{eqnarray}
where we have used the symmetry properties discussed in
the previous section and eq.~(\ref{cona+b}) for the contractions 
$\langle {\mathbf{a^{+} b}}\rangle$.

Using this formula, we determine the phase of the overlap with 
a method similar to the one developed in ref~\cite{HHR82,Ena99}.
Starting from $\phi=0$, for which n($\phi$)=0, 
the value of the phase at neighboring angles is determined by:
\begin{eqnarray}
\Omega(\phi + \delta\phi) \backsimeq \Omega(\phi) + \frac{\pi}{2L}
\left[ \frac{\delta}{\delta\phi} \Omega(\phi) +
\frac{\delta}{\delta\phi} \Omega(\phi+\delta\phi) \right]\; ,
\label{phasnum}
\end{eqnarray}
where we have taken into account that the interval of integration 
has a length $\pi$ and is divided into L segments of equal dimension
$\delta\phi=\pi/L$. 
From the comparison between $\Omega(\phi+\delta\phi)$ calculated 
from (\ref{phasnum}) and $\Phi(\phi +\delta\phi)$ obtained 
from (\ref{phasedolia}), one determines n$(\phi)$, provided the mesh
is dense enough ($\delta\phi$ small enough).

\subsection{Calculation of multipole moments and transition probabilities}

The determination of transition probabilities requires the calculation
of the matrix element of a tensor of order L, $\hat{T}^{M}_{L}$, between
eigenstates of the angular momentum operator.

An eigenstate of the angular momentum operator, with
eigenvalue $J$,  is obtained by projecting the
mean-field wave function $|\Phi\rangle$:
\begin{eqnarray}
|\Phi,JM\rangle = \frac{\sum_K g_K \hat{P}^{J}_{MK} |\Phi\rangle}{
\{\sum_K |g_{K}|^{2} \langle\Phi|\hat{P}^{J}_{KK}|\Phi\rangle\}^{1/2} }\; ,
\end{eqnarray}
where the projector is given by~\cite{RS80}:
\begin{eqnarray}
\hat{P}^{J}_{MK} = \frac{2J+1}{8\pi^{2}} \int d\Omega
{D^{J*}_{MK}}(\Omega) \hat{R} (\Omega)\; ,
\end{eqnarray}
with $\Omega\equiv(\alpha, \beta,\gamma)$ the Euler angles and
$\hat{R}(\Omega) \equiv e^{i\alpha \hat{J}_{z}}e^{i\beta
\hat{J}_{y}}e^{i\gamma \hat{J}_{z}}$.

Only the $K$=0 term of this expression is present
with the symmetry restrictions that we have imposed.
Then:
\begin{eqnarray}
\langle JM,\Phi | \hat{T}^{0}_{L} | J^{'} M^{'}, \Phi^{'} \rangle
=
\frac{ \langle \Phi | \hat{P}^{J}_{0M}
\hat{T}^{0}_{L}\hat{P}^{J'}_{M'0} |\Phi'\rangle}
{ \{  \langle\Phi|\hat{P}^{J}_{00} |\Phi\rangle
  \langle\Phi'|\hat{P}^{J'}_{00} |\Phi'\rangle\}^{1/2} }\;,
\end{eqnarray}
where we have used the properties $(\hat{P}^{J}_{MK})^{+} =
\hat{P}^{J}_{KM}$ and $\hat{P}^{J}_{MK}\hat{P}^{J'}_{M'K'} =
\delta_{JJ'}\delta_{KM'}\hat{P}^{J}_{MK'}$ of the angular momentum
projector operator~\cite{Won75}.

We have
\begin{eqnarray}
&&\langle JM,\Phi | \hat{T}^{0}_{L} | J^{'} M^{'}, \Phi^{'} \rangle =
\frac{\sqrt{(2J+1)(2J'+1)}}{8\pi^{2}} \nonumber \\
&&\frac{\int d\Omega d\Omega' D^{J*}_{0M}(\Omega)D^{J'*}_{M'0}(\Omega')
\langle \Phi| \hat{R}(\Omega)\hat{T}^{0}_{L} \hat{R}(\Omega')
|\Phi'\rangle}
{ \left\{ \left[\int d\Omega D^{J*}_{00}(\Omega) \langle \Phi |
\hat{R}(\Omega)|\Phi\rangle \right]
\left[\int d\Omega' D^{J'*}_{00}(\Omega') \langle \Phi' |
\hat{R}(\Omega')|\Phi'\rangle \right] \right\}^{1/2} }\; .
\end{eqnarray}
The axial symmetry reduces the integration interval to $[0,\pi/2]$.
Moreover, thanks to the transformation
of the wave functions under rotation (see eq.~\ref{paritywf}),
the matrix element
$\langle\Phi|\hat{T}^{0}_{L}\text{e}^{i\beta\hat{J}_{y}}|\Phi'\rangle$
is real and is the same for rotations of angles
$\beta$ and $-\beta$.
In the case of axial symmetry, the final expression takes then the form:
\begin{eqnarray}\label{finalexp}
\langle JM,\Phi| \hat{T}^{0}_{L} |J'M', \Phi'\rangle = \frac{1}{2}
\sqrt{\frac{2J'+1}{2J+1}} \langle J'ML0|JM\rangle \langle
J'0L0|J0\rangle  \delta_{MM'} \nonumber \\
\frac{ \left[ \int \sin\beta d\beta d^{J}_{00}(\beta)
\langle\Phi|\text{e}^{i\beta\hat{J}_{y}} \hat{T}^{0}_{L} |\Phi'\rangle +
\int \sin\beta d\beta d^{J'}_{00}(\beta)
\langle\Phi'|\text{e}^{i\beta\hat{J}_{y}}\hat{T}^{0}_{L}|\Phi\rangle
\right]}
{\left[ \int \sin \beta d\beta d^{J}_{00}(\beta)
\langle\Phi|\text{e}^{i\beta\hat{J}_{y}}|\Phi\rangle\right]^{1/2}
\left[ \int \sin \beta' d\beta' d^{J'}_{00}(\beta')
\langle\Phi'|\text{e}^{i\beta'\hat{J}_{y}}|\Phi'\rangle\right]^{1/2} }
\; .
\end{eqnarray}

In the case of electric quadrupole transitions,
the diagonal matrix element takes the form:
\begin{eqnarray}
&&\langle JM=0,\Phi|\hat{Q}_{20}|JM=0,\Phi\rangle =
\langle J020|J0\rangle^{2}
\frac{ \left[ \int \sin\beta d\beta d^{J}_{00}(\beta)
\langle\Phi|\text{e}^{i\beta\hat{J}_{y}}\hat{Q}_{20}|\Phi\rangle\right]}
{\left[ \int \sin\beta d\beta d^{J}_{00}(\beta)
\langle\Phi|\text{e}^{i\beta\hat{J}_{y}}|\Phi\rangle\right]} \nonumber
\\
&=&  \frac{ (J+1)J } {(2J+3)(2J-1)}
\frac{ \left[ \int \sin\beta d\beta d^{J}_{00}(\beta)
\langle\Phi|\text{e}^{i\beta\hat{J}_{y}}\hat{Q}_{20}|\Phi\rangle\right]}
{\left[ \int \sin\beta d\beta d^{J}_{00}(\beta)
\langle\Phi|\text{e}^{i\beta\hat{J}_{y}}|\Phi\rangle\right]}
\end{eqnarray}
The transition matrix elements
between GCM states are obtained as the weighted sums
of the contributions of the different basis states.

We finally obtain for the reduced transition probability between the
initial, l$_{i}$-th GCM collective state of spin I$_{i}$, to the
final,   l$_{f}$-th GCM collective state of spin I$_{f}$:
\begin{eqnarray}
&&B(EL, I_{i}^{(l_{i})} \rightarrow I_{f}^{(l_{f})} ) = \frac{e^{2}}{4}
\langle I_{i}0L0|I_{f}0\rangle^{2} \nonumber \\
&&\frac{\left[ \int\int da da' f^{(I_{f}^{+},l_{f})*}(a)f^{(I_{i}^{+},l_{i})}(a')
\int dcos\beta d^{I_{f}}_{00}(\beta) \langle \Phi(a) |
\text{e}^{i\beta\hat{J}_{y}} {\mathcal{M}}_{L0} | \Phi'(a') \rangle +
( I_{f}^{+}, l_{f} \leftrightarrow I_{i}^{+}, l_{i})
\right]^{2} }
{ \left[ \int\int da da' f^{(I_{f}^{+},l_{f})*}(a)f^{(I_{f}^{+},l_{f})}(a')
\int dcos\beta d^{I_{f}}_{00}(\beta) \langle \Phi(a) |
\text{e}^{i\beta\hat{J}_{y}}  | \Phi'(a') \rangle \right]
\left[ ( I_{f}^{+}, l_{f} \leftrightarrow I_{i}^{+}, l_{i}) \right] },
\nonumber \\
\end{eqnarray}
with ${\mathcal{M}}_{L0}=r^{2}Y_{L0}(\theta,\phi)$.

\subsection{Density dependence of the effective interactions}
\label{ddofsf}

The density dependent term of the interaction must be
generalized to calculate non diagonal matrix elements.
In the case of a density dependence equivalent to
a 3-body interaction, the Hamiltonians kernel can be
expressed in terms of the left right mixed density~\cite{BDF90}:
\begin{equation}\label{mixdens}
\rho(\mathbf{r})=\sum_{\mu\nu\sigma}
\langle {\mathbf{a}}^{+}{\mathbf{b}} \rangle_{\mu\nu}
\Phi^{*}_{l,\mu}(\mathbf{r},\sigma)\Phi_{r,\nu}(\mathbf{r},\sigma)
\end{equation}
We have chosen the same dependence on the mixed density when there
is no equivalence with a three-body interaction.
The energy is then expressed as a functional of $| R\rangle$ and
$| L\rangle $ similar to the mean-field functional.
One can show that the mixed density depends only on the
relative angles between the principal axes of $| R\rangle$ and
$| L\rangle $. Therefore, after integration on the Euler angles, the
energy is real and does not depend on the orientation of the
reference frame. One can thus restore symmetries either on
the left or the right wave function.

\section{Application to $^{24}$Mg}

 The results shown in this section have been obtained using
the HF+BCS wave functions generated with an axial quadrupole constraint.
The Lipkin-Nogami prescription has been used to improve
the treatment of pairing correlations. It has indeed been shown
that this prescription permits to generate wave functions
which are reasonable approximations of those obtained
by a variation after projection on the good particle number~\cite{Mag93}. 
In this way, the lack of a complete variation
after projection should be partly compensated. The mean-field
results that we will present below correspond to these
HF+BCS+LN calculations.

Since our aim is mainly to test the properties of our method,
we have performed calculations with three Skyrme parameterizations
which have given satisfactory results in the description
of rotational bands in well deformed nuclei, namely
SIII~\cite{Bei74}, SkM$^*$~\cite{Bra82}, and SLy4~\cite{Cha98}.
The pairing interaction is a zero range interaction similar
to the ones used in previous studies of super-deformed bands~\cite{Ter95},
and of nuclei far from stability~\cite{Ter96}. 
In calculations performed with SLy4,
we have varied the strength of the density-dependent pairing force
from  $G= 1250 MeV fm^3$ to $G= 900 MeV fm^3$.
Most results did not depend significantly on the value of G,
the most sensitive quantity being the 
excitation energy of the first 2$^+$ state. 
We have chosen to show here only
results obtained with $G= 1000 MeV fm^3$ which
gives a 2$^+$ energy close to experiment.
The same value of the pairing strength has been
used for the two other Skyrme parametrizations.
It is clear that a precise adjustment of the pairing strength
requires a study of a large range of isotopes,
a work which is in progress.
We have also tested the density independent zero-range
pairing interaction that has been adjusted for
a study of super-heavy nuclei~\cite{Cwi96}.

\begin{figure}[b!] 
 \centerline
 {\epsfig{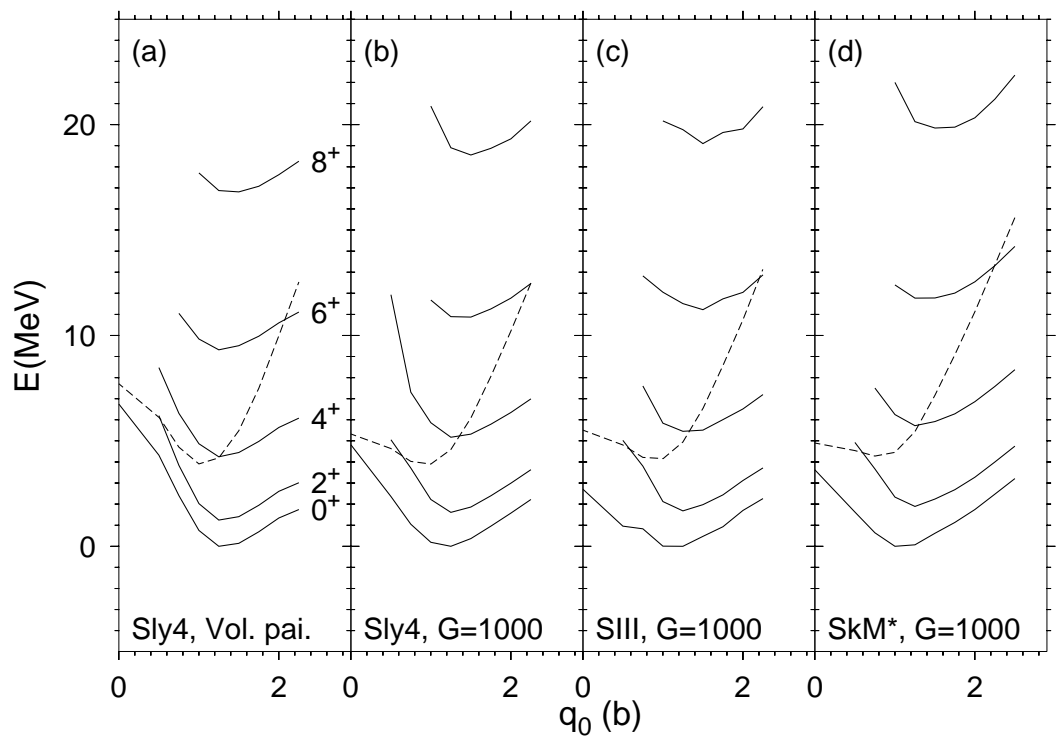}}
\vspace{10pt}
\caption{Mean-field and projected energies obtained  for $^{24}$Mg
as a function of the axial quadrupole
moment q$_{0}$. The pairing interaction is a zero range
interaction without (a) and with (b-d) a density dependence.
The Skyrme parameterizations are SLy4 (a and b), SIII (c)
and SkM$^*$ (d). The dashed lines correspond to
the HF+BCS+LN energies.
The energies obtained by projecting on
angular momentum (from 0$^+$ to 8$^+$) intrinsic wave functions
curves are plotted in full line as a function of the quadrupole
moment of the intrinsic wave function.}
\label{fig1}
\end{figure}

In figure 1 are shown the mean-field energies as a function of
prolate deformations
with the three Skyrme interactions and, in the case of SLy4,
of a surface and a volume pairing force
(dashed curves). Since oblate
deformations play a minor role in this nucleus,
they have not been represented.
The curves are similar, with a well deformed prolate
minimum corresponding to a mass quadrupole moment of
approximately 1 b. The curvature is slightly different
in the four cases, the volume pairing leading to the
deepest minimum.

The energies obtained by projecting each of the mean-field
wave functions on good particle number and angular momentum
are also shown on figure 1. The abscissa of the projected
energies correspond to the quadrupole moment
of the intrinsic wave function.
The spherical configuration is a pure 0$^+$ state and
contributes only to the 0$^+$ projected curve. The energy
gained by projection in this case is due to the difference
between the Lipkin Nogami approximation of the
energy gain due to projection on particle number,
 $-\lambda_2\Delta N^2$, and the exact gain. It is of
the order of 1 MeV in all cases, except in the calculation
with the SIII force, where it is significantly larger.
This may be due to that a lower value for the pairing energy
is obtained in the SIII calculation and that the LN prescription
is less accurate in the low pairing regime.

In all cases, the projection on angular momentum increases
the energy difference between the spherical
configuration and the minimum of the J=0$^+$ curve by 3 to 4 MeV.
The intrinsic wave function leading to the minimum
of this curve has a quadrupole moment slightly larger
than the one corresponding to the mean-field minimum.
For higher angular momenta, the minima are shifted to
higher quadrupole moments. The dependence of the curves
on the nuclear interaction is rather weak
for all J values other than 0; for J=0, the differences are
mainly related to differences between the spherical configurations.

On figure 2, the weights of the various angular
momenta in the mean-field wave functions are plotted
as a function of the axial quadrupole moment. These weights do
not depend significantly on the nuclear interaction;
results obtained with the SLy4
force and a surface pairing are only shown.

\begin{figure}[b!] 
 \centerline{\epsfig{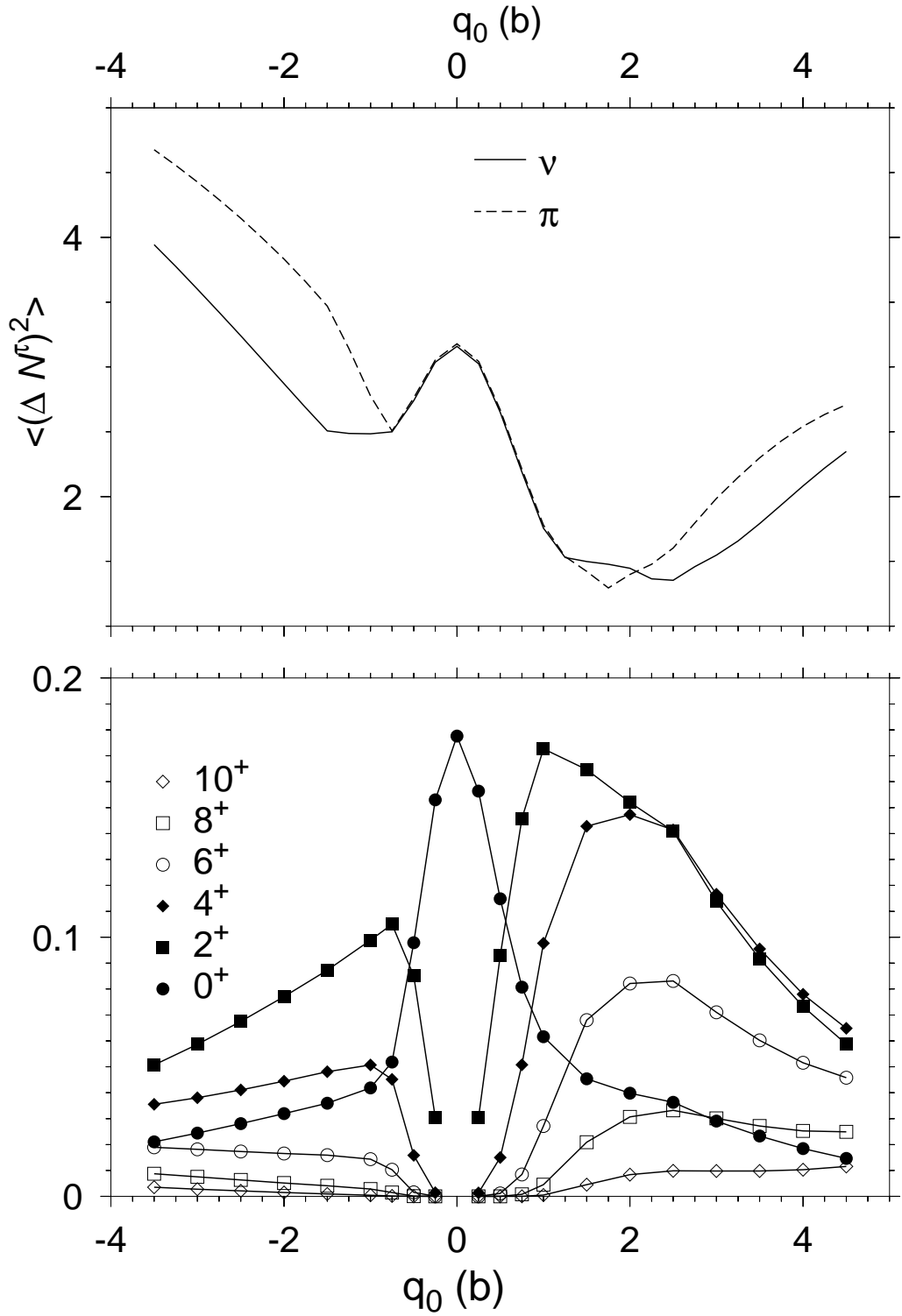}}
\vspace{10pt}
\caption{Lower part: weights $\langle \Psi^{N_{0},Z_{0}}_{J_{0},M=0}
|\Phi(q_{0})\rangle$ as a function of q$_{0}$ and $J$
calculated in case (b) of fig. 1.
Upper part: particle number dispersions
in the intrinsic wave functions as a function of $q_{0}$
for neutrons (full line) and for protons (dashed line).}
\label{fig2}
\end{figure}

The spherical configuration is
already an eigenstate of the angular momentum
operator and is
affected only by the particle number projection.
Its N=Z=12 component has a weight of 0.18, the remaining part 
corresponding to different proton and neutron numbers. 
The J=0 curve displays a
pronounced peak around the spherical
point. It is not symmetric with respect to
quadrupole moment, the weights
on the prolate side being larger than on the oblate side.
For this reason, the J=0 curve is rather far from a Gaussian.
This shows that the Gaussian overlap approximation
sometimes used to derive a collective Shrodinger equation
should be used with care.
When the weight of a mean-field
wave function  component is lower than 0.01, the projection becomes
numerically difficult; such tiny components have been
excluded from the configuration mixing calculations.
As expected, the quadrupole moment correspondig to
the largest weight moves to
higher quadrupole moments as a function of angular
momentum. For J=10$^+$, no maximum is obtained up to
a quadrupole moment of 4.5 b, showing that
our variational space is inadequate to
describe accurately states with
such a high spin. A dissymetry between the
prolate and oblate deformations can also be noticed
on figure 2: the total weights corresponding to N=Z=12
are twice larger for prolate configurations than for oblate ones.
This confirms that $^{24}$Mg is dominated by prolate configurations.
The different dispersions in particle numbers for
prolate and oblate deformations is at the
origin of this dissymetry (upper part of the figure).

\begin{figure}[b!] 
 \centerline{\epsfig{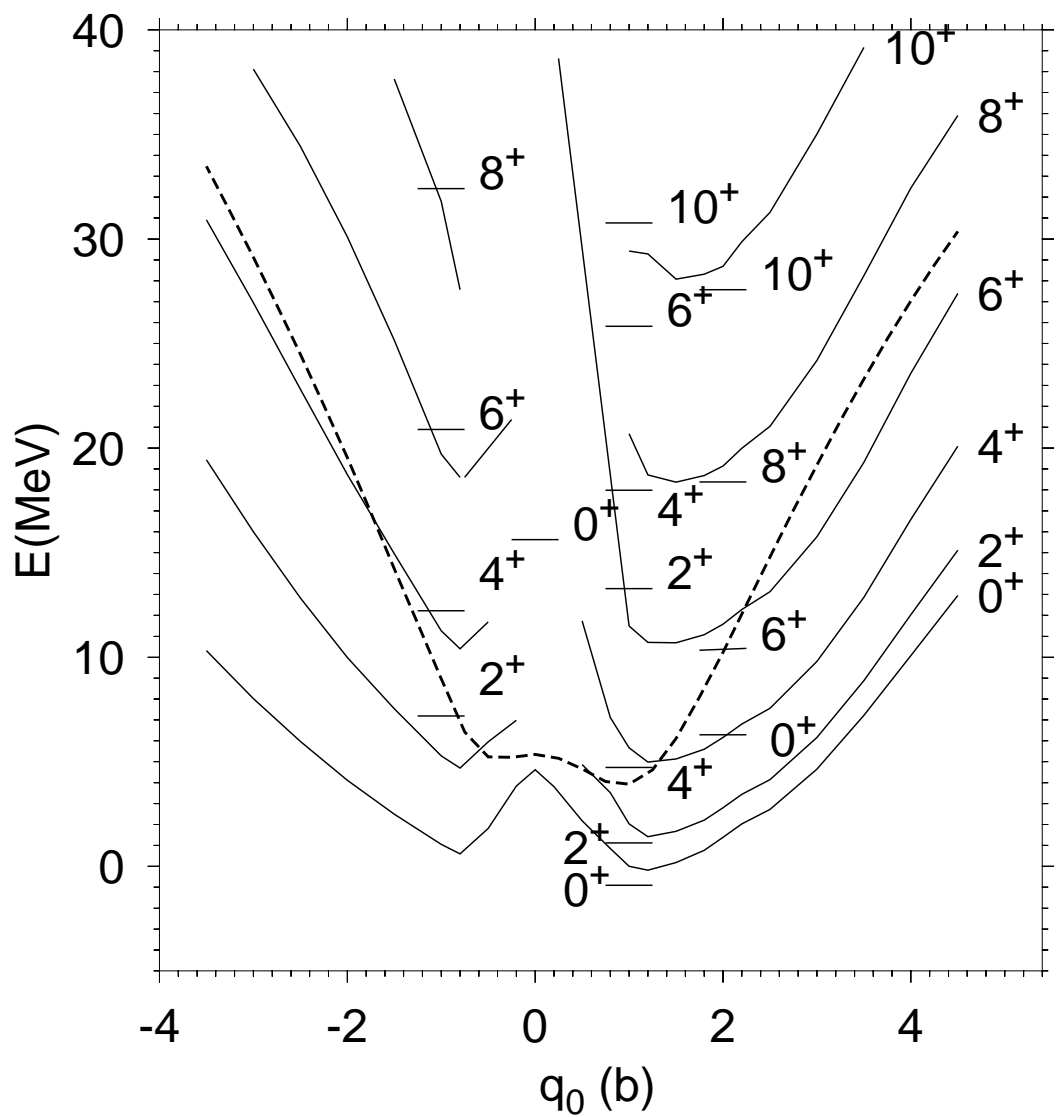}}
\caption{Projected energies for $^{24}$Mg as a function of the axial
quadrupole moment (case (b) of fig. 1).
The symbols are the same as in the previous
figures. The first three energies obtained for each angular momentum
in the configuration mixing calculation
are represented by horizontal bars centered at the value of q$_{0}$
where the respective collective wave functions are maximum.}
\label{fig3}
\end{figure}

The variation of the energy as a function of prolate and
oblate deformations
is plotted on figure 3 for the SLy4 Skyrme parametrization
and a surface pairing interaction.
In addition to the prolate minimum already observed
on figure 1, the mean-field curve presents a shoulder
at an  oblate deformation around 0.5 b.
The full projection creates an oblate minimum at the
position of that shoulder for J values 
ranging from 0$^+$ to 6$^+$.
For greater values of $J$,
the weights of the intrinsic wave functions
 for deformations below --2\,b are very small.
Consequently, the projected energy curves
do not exhibit any oblate minima.
However the J=0$^+$ to 6$^+$ minima are probably not stable
against triaxial deformations, since a calculation
including triaxial deformations indicates that
the shoulder in the intrinsic curve is a maximum
with respect to $\gamma$.

\begin{figure}[b!] 
 \centerline{\epsfig{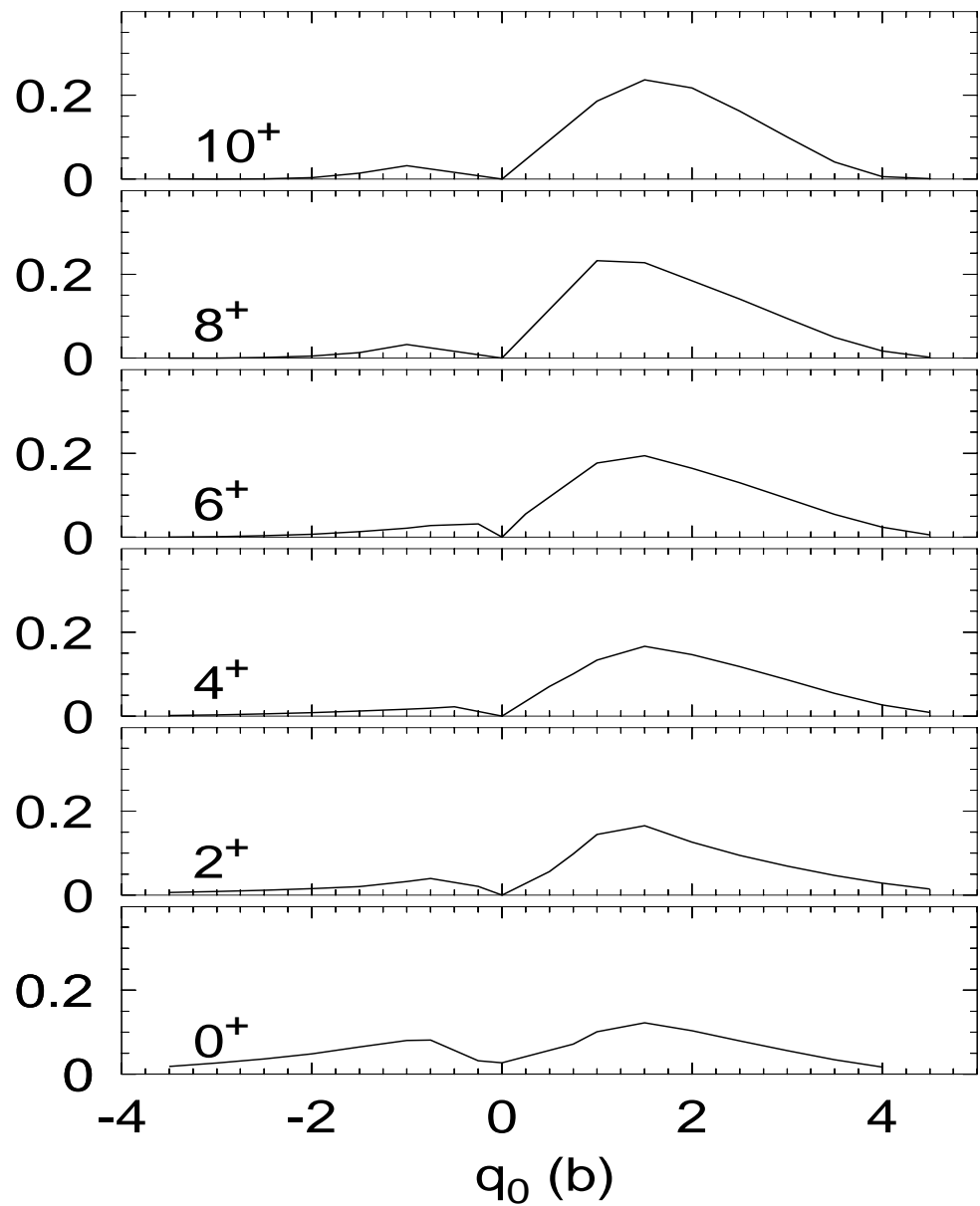}}
\vspace{10pt}
\caption{Squared weights  of intrinsic
states in the Yrast levels ($l=0$) of
the different angular momentum representations. }
\label{fig4}
\end{figure}

For each value of the angular momentum, we have performed a
configuration mixing calculation including quadrupole
moments between --3.5 b and 4.5 b.
This corresponds to intrinsic configurations excited by
about 30\,MeV with respect to the prolate minimum.
This configuration mixing is nothing but a variation
after projection in a limited but hopefully relevant
Hilbert space.
The spectrum 
generated in this way (represented by bars) is plotted
at the quadrupole moment corresponding to the largest
component of the collective wave function,
which are shown on figure 4 for the yrast states.
The value
of this quadrupole moment
is very close to the minimum of the projected
energy curve.
Moreover, the energy of this minimum is slightly modified
by the configuration mixing. The largest gain,
$\sim$800\,keV, is obtained for the 0$^+$ state,
but is reduced at higher spins.
Several excited states are found at low energy for each spin value.
Except for the second 0$^+$ and 10$^+$, the wave functions of yrare
states are peaked around the oblate secondary minimum.

\begin{figure}[b!] 
 \centerline{\epsfig{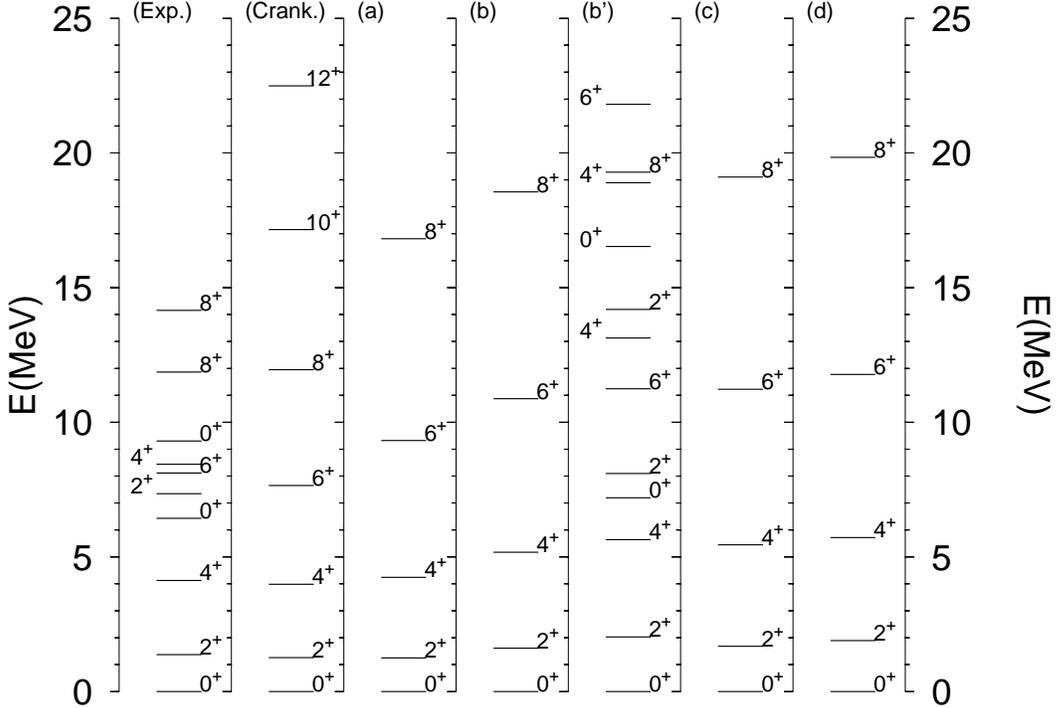}}
\vspace{10pt}
\caption{Experimental positive parity levels of $^{24}$Mg
compared to a cranking calculation (SLy4 and density-dependent
zero-range pairing interactions) and to the (a)-(d) choices of interactions
described in fig.1. In the column (b') is plotted the configuration
mixing spectrum.}
\label{fig5}
\end{figure}

The spectra obtained with the four choices of interactions
described for fig. 1 are compared to the experimental
spectrum~\cite{Pri86} on figure 5.
The energies correspond to the minima of the projected
energy curves except in the column (b') where
the configuration mixing spectrum is represented.
For comparison, we show
also the results of a cranking
calculation performed with the SLy4 interaction and
a surface pairing with the same strength of 1250MeV
that we used in our calculations of SD rotational bands.
In this case, we use the HFB+Lipkin-Nogami method
presented in our previous cranking calculations~\cite{Ter95}.

Since the intrinsic ground state of $^{24}$Mg has
a large prolate deformation, one expects the
cranking approximation to be valid. The cranking spectrum is
indeed in good agreement with the experimental data
up to the 6$^+$ level. Experimentally, it is the second 8$^+$
which belongs to the ground state rotational band and
our result underestimates its energy by 2.0 MeV. The energy
at which we obtain a 10$^+$ state is probably also
too low. In the cranking calculation, the pairing
energy becomes very small at spin 8; triaxiality
effects become also important around this spin.

 In the projected spectra, the energy of
the first 2$^+$ state is systematically overestimated,
the energy obtained with a volume dependent pairing
being however rather close to the experimental value.
Two restrictions imposed in the present calculation
may be at the origin of this discrepancy.
First, we have not included  triaxial deformations,
which are shown by the cranking calculation to be more important
at high spins than at low spins. Second, the  variation
after projection on angular momentum is limited
to the quadrupole moment; the mean-field equations
are optimized for the description of the ground state energy
and not for excited states. The first limitation could be removed
with minor modifications of the formulae presented in section 2.
However, the computing time would be largely increased, the
number of Euler angles for the angular integration 
being at least 50 times larger.
A rather simple way to improve the variational character
of the calculation would be to project for each spin
wave functions generated by cranking calculations. It
is indeed well known~\cite{RS80} that  cranking is a first order approximation
of a variation after projection on angular momentum. To generalize
our method in this direction requires mainly to consider Bogoliubov
instead of BCS   transformations, a fact which would not increase
too much the computing time. Work along this line is in progress.

\begin{figure}[b!] 
 \centerline{\epsfig{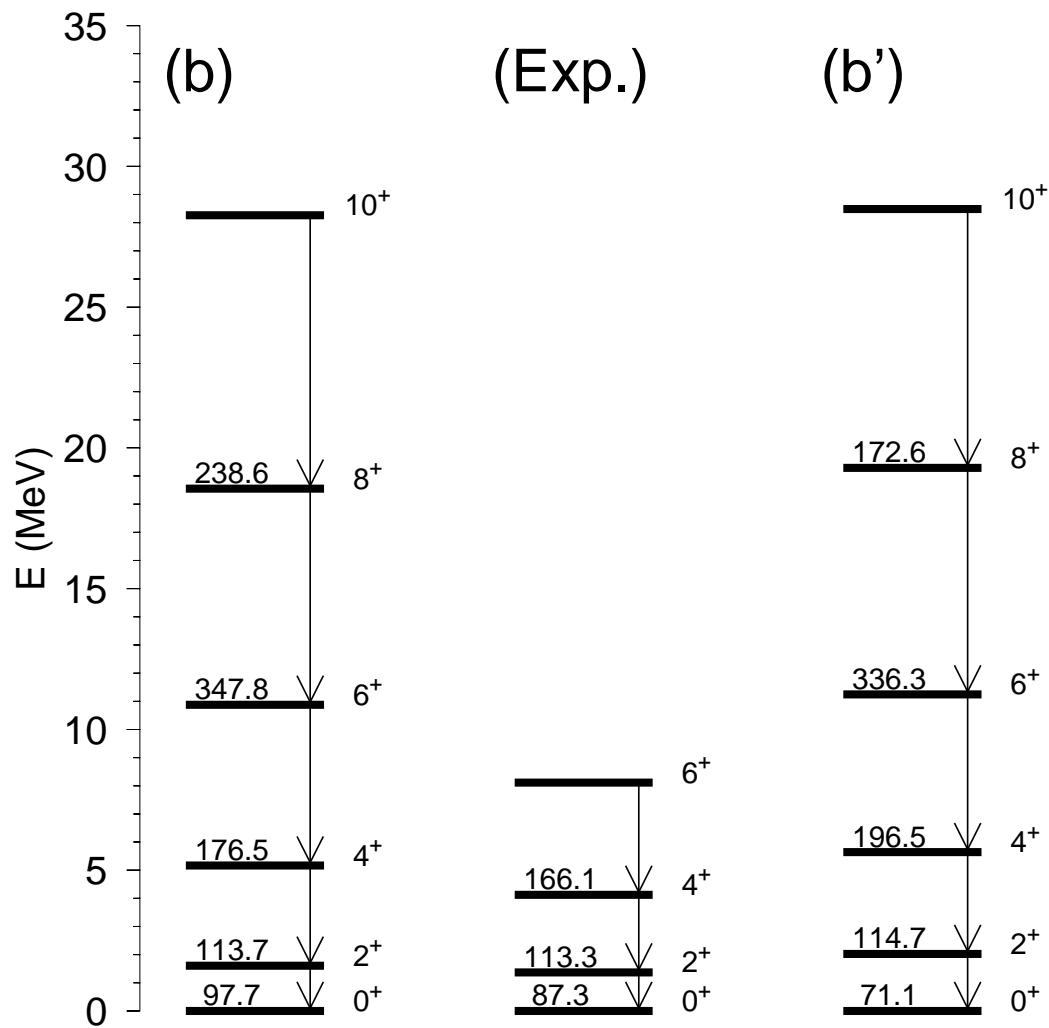}}
\vspace{10pt}
\caption{$B(E2)$ transition probabilities (in~$e^2fm^4$ for $^{24}$Mg.
Transition probabilities between the configurations corresponding
to the  minimum of the
projected energy curves of figure 3 (left side) and between the yrast collective
states obtained in the GCM calculation (right side).
In the central part are shown
the experimental values~\cite{Pri86}.}
\label{fig6}
\end{figure}

One of the main interests of a restoration of rotational symmetry
is the possibility to calculate transition probabilities
without the approximation involved in a cranking calculation.
On figure 6, are compared to the experimental data~\cite{Pri86} the
transition probabilities along the yrast line
obtained in the GCM calculation and by considering only
the minima of projected energy curves.
The value of the 2$^+$ to 0$^+$ B(E2)
is nearly independent of the nuclear interaction
and results are only shown for the SLy4 interaction (case b).
The transition probability between the configurations
minimizing the projected energy curves is very close to the
experimental value. The configuration mixing causes
a spreading of the collective wave function on the quadrupole moment
and decreases slightly the value of the B(E2).
This effect is similar to the effect of quadrupole vibrations
that is sometimes included phenomenologically~\cite{Rei99} in the determination
of transition probabilities from intrinsic wave functions.
Since for spin different from 0, the wave functions do not
have components at low quadrupole moment, the configuration mixing
does not affect significantly the transition probabilities.
The agreement between both calculations and the experimental
data is excellent in these cases.

\begin{figure}[b!] 
 \centerline{\epsfig{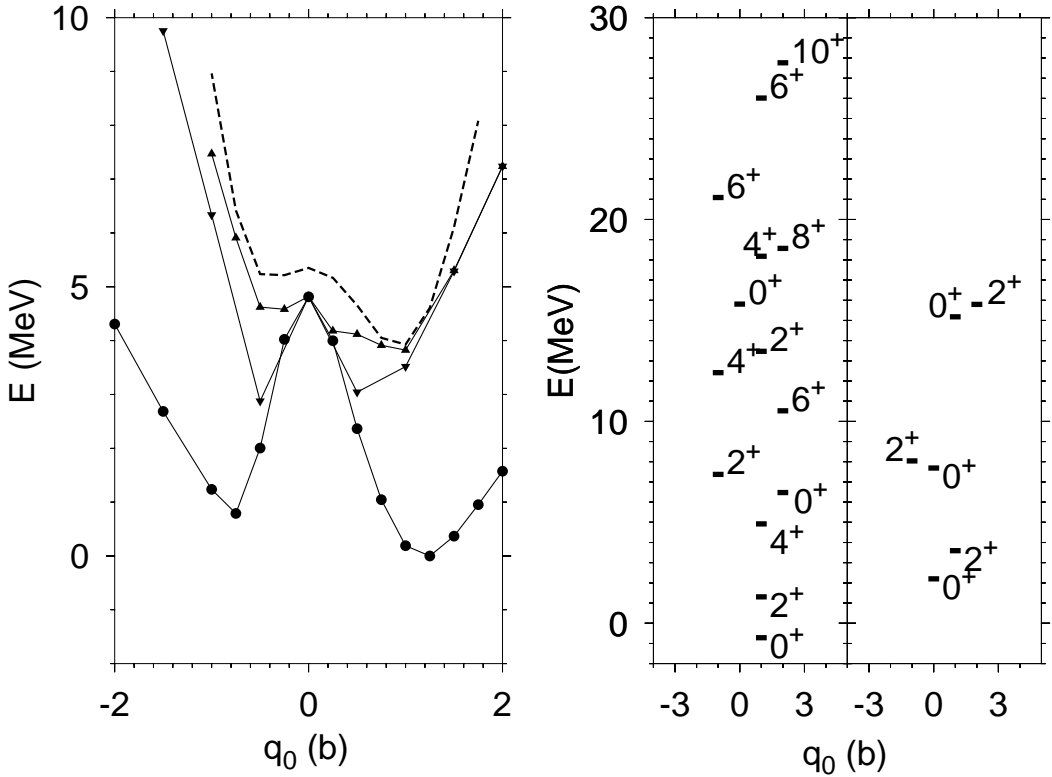}}
\vspace{10pt}
\caption{On the left part of the figure are shown
the energy curves corresponding to the mean-field calculation
(dashed line),
projection on particle number only (up triangles),
projection on particle number
together with mixing of the three possible orientations of the
principal axes (down triangles) and triple projection (dots).
On the right are compared the configuration mixing energies
obtained with the full (on the left) and the restricted
(on the right) projections.
}
\label{fig7}
\end{figure}

In our first studies of quadrupole collective dynamics~\cite{BDF91},
we have introduced an approximate projection on 0$^+$ and 2$^+$
states from intrinsic states. This projection is based on the mixing
of the configurations corresponding to all the possible labellings
of the principal axes of inertia. It is supposed to be valid for small
deformations. The present calculation gives the opportunity
to test the validity of this approximate projection. On figure 7
are plotted the energy curves corresponding to the mean-field calculation,
projection on particle number only, projection on particle number
together with mixing of the three possible orientations of the
principal axes and  triple projection.
The effect of particle number projection is dominant for small
deformations, corresponding to a $\beta$ value up to 0.1.
The restricted projection on angular momentum
gives a fair approximation for slightly larger deformations,
up to $\beta$ around 0.2. For still larger
quadrupole moments, the full projection becomes necessary.
On the right part of the figure are compared the configuration
mixing spectra obtained with the exact and the approximate
projections.  The differences in  total energy
are due to the failure of the approximate projection
above quadrupole moments of 0.5 b. However,
the relative position of the first 0$^+$ and 2$^+$
is rather satisfactorily estimated.

The restricted projection is of limited
interest for a well deformed nucleus like $^{24}$Mg.
However for heavy spherical nuclei,
its validity up to a $\beta$ value  around 0.2
makes it a cheap alternative to study quadrupole dynamics.

\section{Conclusion}

In this paper, we have presented and tested a method to introduce
correlations beyond mean-field on HF+BCS wave functions.
The formulae given in the first part of section 2
are written in a form appropriate for the relaxation
of the restrictions imposed in this study.
The extensive tests performed on the $^{24}$Mg nucleus show 
that the method works with reasonable computing time.

A first and natural generalization from BCS 
to full Bogoliubov-Valatin transformations is under progress. 
It will allow a better treatment for pairing correlations.
If no significant improvements for even-even nuclei close to the
stability line is to be expected, HFB is essential to treat 
correctly nuclei near the drip lines (see ref.~\cite{Ter96,Dob84}).
 
As the projection onto angular momentum implies 
that one cannot anymore make use of the signature symmetry,
the generalization to many-body wave functions breaking time
reversal invariance is a natural next step.
That is a necessary step towards a 
description of odd nuclei. It will also enable to
project wave functions generated for each spin
by cranking calculations. As it has already been 
shown theoretically~\cite{Man75,RS80},
the use of cranking wave functions is the first
order of a variation after projection on angular
momentum. Numerical applications~\cite{Bay84,HHR82}
have confirmed that the projection of cranking wave functions
improves the energy obtained for each angular momentum
and compresses the spectra. Such an effect would correct
the too spread spectra obtained in the present study.

 Another important question for which we have preliminary
answers concerns the use of effective interactions
adjusted for mean-field calculations in a model
where are incorporated correlations beyond HF+BCS. 
The Skyrme parametrizations that we have
tested here behave in a similar way and are reasonable 
starting points
to study the effects of correlations. 
We have started a study of
several Mg isotopes and of a few neighbouring nuclei 
is underway. This should enable to better determine
the properties of effective interactions on which 
depend the spectra of nuclei far from stability.

\section {Acknowledgements}      
 This research was supported in part by
the PAI-P3-043 of the Belgian
Office for Scientific Policy.

\end{document}